\documentclass[twocolumn]{aastex61}
\pdfoutput=1 
\usepackage{amsmath,amstext}
\usepackage[T1]{fontenc}
\usepackage{apjfonts} 
\usepackage[figure,figure*]{hypcap}

\newcommand{\HST}{\textit{HST}}
\shorttitle{Environment of a neutron star merger}
\shortauthors{Levan et al.}

\begin{document}

\title{The environment of the binary neutron star merger GW170817}
\author{A.J. Levan}
\affiliation{Department of Physics, University of Warwick, Coventry, CV4 7AL, UK}
\author{J.D. Lyman}
\affiliation{Department of Physics, University of Warwick, Coventry, CV4 7AL, UK}
\author{N.R. Tanvir}
\affiliation{Department of Physics and Astronomy, University of Leicester, LE1 7RH, UK}
\author{J. Hjorth}
\affiliation{Dark Cosmology Centre, Niels Bohr Institute, University of Copenhagen, Juliane Maries Vej 30, Copenhagen \O, 2100, Denmark}
\author{I. Mandel}
\affiliation{Birmingham Institute for Gravitational Wave Astronomy and School of Physics and Astronomy, University of Birmingham, Birmingham, B15 2TT, UK}
\author{E.R. Stanway}
\affiliation{Department of Physics, University of Warwick, Coventry, CV4 7AL, UK}
\author{D. Steeghs}
\affiliation{Department of Physics, University of Warwick, Coventry, CV4 7AL, UK}
\author{A.S. Fruchter}
\affiliation{Space Telescope Science Institute, 3700 San Martin Drive, Baltimore, MD 21218, USA}

\author{E. Troja}
\affiliation{Department of Astronomy, University of Maryland, College Park, Maryland 20742-4111, USA}
\affiliation{Astrophysics Science Division, NASA Goddard Space Flight Center, 8800 Greenbelt Rd, Greenbelt, MD 20771, USA}

\author{S. L. Schr\o der}
\affiliation{Dark Cosmology Centre, Niels Bohr Institute, University of Copenhagen, Juliane Maries Vej 30, Copenhagen \O, 2100, Denmark}

\author{K. Wiersema}
\affiliation{Department of Physics and Astronomy, University of Leicester, LE1 7RH, UK}

\author{S. H. Bruun}
\affiliation{Dark Cosmology Centre, Niels Bohr Institute, University of Copenhagen, Juliane Maries Vej 30, Copenhagen \O, 2100, Denmark}

\author{Z. Cano}
\affiliation{Instituto de Astrof\'isica de Andaluc\'ia (IAA-CSIC), Glorieta de la Astronom\'ia, s/n, 18008, Granada, Spain}

\author{S.B. Cenko}
\affiliation{Joint Space-Science Institute, University of Maryland, College Park, MD 20742, USA }
\affiliation{Astrophysics Science Division, NASA Goddard Space Flight Center, 8800 Greenbelt Rd, Greenbelt, MD 20771, USA}

\author{A. de Ugarte Postigo} 
\affiliation{Instituto de Astrof\'isica de Andaluc\'ia (IAA-CSIC), Glorieta de la Astronom\'ia, s/n, 18008, Granada, Spain}
\affiliation{Dark Cosmology Centre, Niels Bohr Institute, University of Copenhagen, Juliane Maries Vej 30, Copenhagen \O, 2100, Denmark}

\author{P. Evans}
\affiliation{Department of Physics and Astronomy, University of Leicester, LE1 7RH, UK}

\author{S. Fairhurst}
\affiliation{School of Physics and Astronomy, Cardiff University, Cardiff, United Kingdom,
CF24 3AA}

\author{O.D. Fox}
\affiliation{Space Telescope Science Institute, 3700 San Martin Drive, Baltimore, MD 21218, USA}

\author{J.P.U. Fynbo}
\affiliation{Dark Cosmology Centre, Niels Bohr Institute, University of Copenhagen, Juliane Maries Vej 30, Copenhagen \O, 2100, Denmark}

\author{B. Gompertz}
\affiliation{Department of Physics, University of Warwick, Coventry, CV4 7AL, UK}

\author{J. Greiner}
\affiliation{Max-Planck-Institut f\"{u}r extraterrestrische Physik, 85740 Garching, Giessenbachstr. 1, Germany}

\author{M. Im}
\affiliation{Center for the Exploration of the Origin of the universe (CEOU), Seoul National University, Seoul, Korea; Astronomy Program, Department of Physics \& Astronomy, Seoul National University, Seoul, Korea}

\author{L. Izzo}
\affiliation{Instituto de Astrof\'isica de Andaluc\'ia (IAA-CSIC), Glorieta de la Astronom\'ia, s/n, 18008, Granada, Spain}

\author{P. Jakobsson}
\affiliation{Centre for Astrophysics and Cosmology, Science Institute, University of Iceland, Dunhagi 5, 
107 Reykjav\'ik, Iceland}

\author{T. Kangas}
\affiliation{Space Telescope Science Institute, 3700 San Martin Drive, Baltimore, MD 21218, USA}

\author{H.G. Khandrika}
\affiliation{Space Telescope Science Institute, 3700 San Martin Drive, Baltimore, MD 21218, USA}

\author{A.Y. Lien}
\affiliation{Department of Physics, University of Maryland, Baltimore County, 1000 Hilltop Circle, Baltimore, MD 21250, USA}
\affiliation{Astrophysics Science Division, NASA Goddard Space Flight Center, 8800 Greenbelt Rd, Greenbelt, MD 20771, USA}

\author{D. Malesani}
\affiliation{Dark Cosmology Centre, Niels Bohr Institute, University of Copenhagen, Juliane Maries Vej 30, Copenhagen \O, 2100, Denmark}

\author{P. O'Brien}
\affiliation{Department of Physics and Astronomy, University of Leicester, LE1 7RH, UK}

\author{J.P. Osborne}
\affiliation{Department of Physics and Astronomy, University of Leicester, LE1 7RH, UK}

\author{E. Palazzi}
\affiliation{INAF, Institute of Space Astrophysics and Cosmic Physics, Via Gobetti 101, I-40129 Bologna,
Italy}

\author{E. Pian}
\affiliation{INAF, Institute of Space Astrophysics and Cosmic Physics, Via Gobetti 101, I-40129 Bologna,
Italy}

\author{D.A. Perley}
\affiliation{Astrophysics Research Institute, Liverpool John Moores University, IC2, Liverpool Science Park, 146 Brownlow Hill, Liverpool L3 5RF}

\author{S. Rosswog}
\affiliation{The Oskar Klein Centre, Department of Astronomy, AlbaNova, Stockholm University, SE-106 91 Stockholm, Sweden}

\author{R.E. Ryan}
\affiliation{Space Telescope Science Institute, 3700 San Martin Drive, Baltimore, MD 21218, USA}

\author{S. Schulze}
\affiliation{Department of Particle Physics and Astrophysics, Weizmann
Institute of Science, Rehovot 761000, Israel.}

\author{P. Sutton}
\affiliation{School of Physics and Astronomy, Cardiff University, Cardiff, United Kingdom,
CF24 3AA}

\author{C.C. Th\"one}
\affiliation{Instituto de Astrof\'isica de Andaluc\'ia (IAA-CSIC), Glorieta de la Astronom\'ia, s/n, 18008, Granada, Spain}

\author{D.J. Watson}
\affiliation{Dark Cosmology Centre, Niels Bohr Institute, University of Copenhagen, Juliane Maries Vej 30, Copenhagen \O, 2100, Denmark}

\author{R.A.M.J. Wijers}
\affiliation{Anton Pannekoek Institute for Astronomy, University of Amsterdam, Postbus 94249, NL-1090 GE Amsterdam, the Netherlands}

\begin{abstract}
We present {\em Hubble Space Telescope} and {\em Chandra} imaging, combined with Very Large Telescope MUSE integral field spectroscopy of the counterpart and host galaxy of
the first binary neutron star merger detected via gravitational wave emission by LIGO \& Virgo, GW170817. The host galaxy, NGC\,4993, is an S0 galaxy at $z=0.009783$. 
There is
evidence for large, face-on spiral shells in continuum imaging, 
and edge-on spiral features visible in
nebular emission lines. This suggests that NGC 4993 has undergone a relatively recent ($\lesssim 1$ Gyr) ``dry'' merger. 
This merger may provide the fuel for a weak active nucleus seen in {\em Chandra} imaging. At the location of the counterpart, \HST\ imaging implies there is no globular or young stellar cluster, with a limit of a few thousand solar masses for any young system. The population in the vicinity is predominantly old with $\lesssim 1$\% of any light arising from a population with ages $<500$~Myr. Both the host galaxy properties and those of the transient location are consistent with the distributions seen for short-duration gamma-ray bursts, although the source position lies well within the effective radius ($r_e \sim 3$~kpc), providing an $r_e$-normalized offset that is closer than $\sim 90\%$ of short GRBs. For the
long delay time implied by the stellar population, this suggests that the kick velocity was significantly less than the galaxy escape velocity. We do not see any narrow host galaxy interstellar medium features within the counterpart spectrum, implying low extinction, and that the binary may lie in front of the bulk of the host galaxy.
\end{abstract}

\keywords{stars:neutron --- galaxies: individual (NGC 4993) --- galaxies: kinematics and dynamics}

\section{Introduction}
The existence of binary neutron stars that will eventually merge via the loss of angular momentum and
energy through gravitational wave (GW) emission has been recognized since the identification
of the Hulse-Taylor pulsar \citep{hulse75}. These mergers have long been thought to manifest themselves
as short-duration gamma-ray bursts \citep[SGRBs,][]{eichler89}, and may produce additional optical/IR emission
due to the synthesis of radioactive elements in their ejecta \citep[e.g.][]{li98,metzger12,barnes13}. However, direct observations
of confirmed neutron star mergers are challenging, because smoking guns to their nature
have been difficult to come by, and only in few cases have both signatures been reported \citep[e.g.][]{tanvir13,berger13}. 

This has changed with the discovery of GW170817, an
unambiguous neutron star merger directly measured in gravitational waves \citep{lvcdisc17}, associated with
a short duration gamma-ray burst \citep{lvc_fermi} as well as a radioactively-powered kilonova \citep[e.g.][]{pian17,tanvir17,lvccapstone}. For the first time this
provides a route for studying the properties of a confirmed neutron star binary merger in detail. 
In this paper, we consider the environment of the merger, and the constraints this places on the properties of the progenitor binary. 

\section{Observations}
GW170817 was detected by the Advanced LIGO-Virgo observatory network on 2017-08-17:12:41:04 UT \citep{lvc21509}, and has a chirp
consistent with a binary neutron star merger. 
Approximately two seconds later
the {\em Fermi} Gamma-ray Burst Monitor (GBM) triggered on GRB\,170817A \citep{gcn21506,gcn21520,fermi_sub}, a short GRB (duration $\sim 2$s)
that was also seen by {\em INTEGRAL} \citep{gcn21507}. While the sky localisations of both events were large, they overlapped, and the
combined spatial and temporal coincidence suggested causal association \citep{lvc_fermi}. Numerous groups undertook searches of the resulting GW-error region, 
revealing a counterpart in NGC 4993 \citep{coulter17,coulter17paper}, independently confirmed by several groups \citep{valenti17,tanvir17vista,master17,allam17,arcavi17}. 
The counterpart, known as SSS17a/AT~2017gfo, was seen to brighten in the IR and then dramatically redden in the following nights 
\citep{evans17,pian17,smartt17,tanvir17}, revealing broad features consistent with the expectations 
for a transient driven by heavy element (r-process) nucleosynthesis, often dubbed a kilonova \citep[][]{li98,metzger12,barnes13}.
These properties cement the association of the optical counterpart with both the GRB and the gravitational wave trigger. 

We obtained several epochs of ground and space based observations of the counterpart of GW170817. 
Observations with the Very Large Telescope using the MUSE integral field spectrograph were obtained on 2017-08-18. MUSE has a field
of view of 1\arcmin\ and covers the spectral range from 4800 to 9300\,\AA{}. Data were reduced using the ESO pipeline (v2.0.3) via {\tt Reflex}, and fit for stellar continua and emission lines with {\sc ifuanal} as described in \cite{lyman17}. Hubble Space Telescope ({\em HST})
observations were obtained between 2017-08-22
and 2017-08-28
in the F275W, F475W, F606W, F814W, F110W and F160W filters
via programmes GO 14804 (Levan), 14771 (Tanvir) and 14850 (Troja). A description of time variability of the counterpart in these images is provided in \cite{tanvir17}. In addition, a single ACS/F606W observation
of NGC 4993 was taken on 28 April 2017, prior to the discovery. 
Imaging observations were reduced via {\tt astrodrizzle}, with the final scale set to 0.025\arcsec (UVIS) and 0.07\arcsec (IR). In addition we analyze optical images ($u$, 2400~s; $R$, 240~s; $z$, 240~s) of 
NGC 4993 obtained at ESO with the Visible wide field Imager and Multi-Object Spectrograph (VIMOS), on 22 Aug 2017, reduced via the standard {\tt esorex} pipeline, and archival {\em Spitzer} observations, for which we used the processed, post basic calibration (PBCD) data. A log of our observations, and their times is given in \cite{tanvir17}.

In addition, we also present {\em Chandra} observations of the host galaxy. A total of 47~ks were obtained on 01 Sept 2017 (program 18508587, PI Troja). For analysis below we use a cleaned and extracted 0.5--8 keV image. Full details of the {\em Chandra} observations are given in \cite{troja}

\section{The host galaxy at large}
\subsection{Morphology and dynamics}
At first sight, NGC 4993 appears to be a typical S0 galaxy: it has a strong bulge component and some visible dust lanes close to the galaxy core in the {\em HST} imaging (Figure 1), suggestive of recent merger activity in an ancient population. It has a measured redshift from our MUSE data of $z=0.009783$, corresponding to a distance of 42.5 Mpc assuming a Hubble expansion with $H_0 = 69.6$ km s$^{-1}$ and neglecting any peculiar velocity. Based, on photometry in 1\arcmin~ apertures, it has an absolute $K$-band magnitude of $M_K \sim -21.5$\,(AB). A S\'ersic fit to R-band and F606W images
yields an effective radius 
of $\sim 16 \pm 1 \arcsec \approx 3$ kpc, with a S\'ersic index of $n\sim$ 4 that is indicative of a bulge/spheroid dominated galaxy.
A fit to the global spectral energy distribution of the galaxy (see Table 1) 
suggests a stellar mass of 
$M_* \sim 1.4 \times 10^{11}$ M$_{\odot}$ based
on the stellar population models of \cite{maraston05}, 
and little to no ongoing star formation. These diagnostics are typically the only ones available for short-GRB hosts, and indeed the properties of NGC4993 are broadly in keeping with those of the fraction of massive early type 
galaxies that host short GRBs \citep{fong13,fong16}. However, NGC 4993 is much closer than the host galaxies of all previously known short GRBs, making it possible to dissect it in greater detail, in particular with regard to its resolved morphology and the nature of the stellar population(s). 

\subsection{Stellar populations}
\label{sec:sp}
Stellar populations were fit to spaxel\footnote{A spaxel is a spectral pixel in an integral field spectrograph, where each spatial pixel provides spectral coverage} bins across the host using {\sc starlight} \citep{starlight} following the method detailed in \citet{lyman17}. These provide spatially resolved maps\footnote{Spaxel bins used a Voronoi binning algorithm to achieve a minimum SNR of 25 \citep{cappellari03} and remain $\lesssim 1$\arcsec\ in radius, even in the fainter outskirts of the host.} of stellar velocity, velocity dispersion, extinction and ages for NGC~4993. Figure~\ref{musesp} shows the results of our fits. The galaxy is dominated by an older population with $>60\%$ of the mass arising from stars $\gtrsim 5$~Gyr in age in essentially all of the bins. In general only 1-2\% of the light (and thus $\ll 1\%$ of the mass) in the best fits arises from stars with ages $<500$ Myr. The strong Balmer absorption and lack of evidence for young stellar populations is reminiscent of post-starburst or post-merger galaxy spectra. This is borne out in our fits, which indicate a strong contribution from an intermediate ($\sim$~Gyr) stellar population that may be responsible for the ionised gas we see (see Sec 3.3 \& Figure~\ref{musen2}) via post-AGB stars.

For galaxies such as this, it is unsurprising that a single population does not provide a good fit to the resulting data. However, including large numbers of model population ages in the fit, that contribute progressively less light, risks over-interpreting the data or systematics within it and are increasingly prone to degenerate fits. We use a relatively sparse number of model ages (13) and find that repeat fitting with just six ages gave similar results. The contribution of the young stellar population disappears in a large fraction of the bins using the reduced model set, indicating that they contribute no more than 1-2\% of the light.

\subsection{Evidence for past merger and emission line properties}
{\em Chandra} X-ray observations reveal a compact source consistent ($\pm 0.5$\arcsec) with the nucleus of the galaxy, with a luminosity (for a photon index of $\Gamma=2$) of $L_X \approx 2 \times 10^{39}$ erg s$^{-1}$ (0.5 - 8 keV). There is no obvious extended emission associated with features seen at other wavelengths, although further point sources in the vicinity may be associated with NGC 4993. This X-ray emission is most likely due to a weak AGN, in keeping with a 0.4 mJy radio detection at 19 GHz \citep{troja}.
At large scales the velocity dispersion within the
effective radius within our MUSE cube is $\sim 170$ km s$^{-1}$, and this yields a black hole mass from $M_{BH} - \sigma$ relations \cite{ms1,gul09} of
$M_{BH} \approx 10^8$ M$_{\odot}$. {For this black hole mass the AGN is only accreting at $\sim 10^{-6}$ of the Eddington luminosity.}

\begin{figure*}[h]
    \centering
    \includegraphics[width=15cm,angle=0]{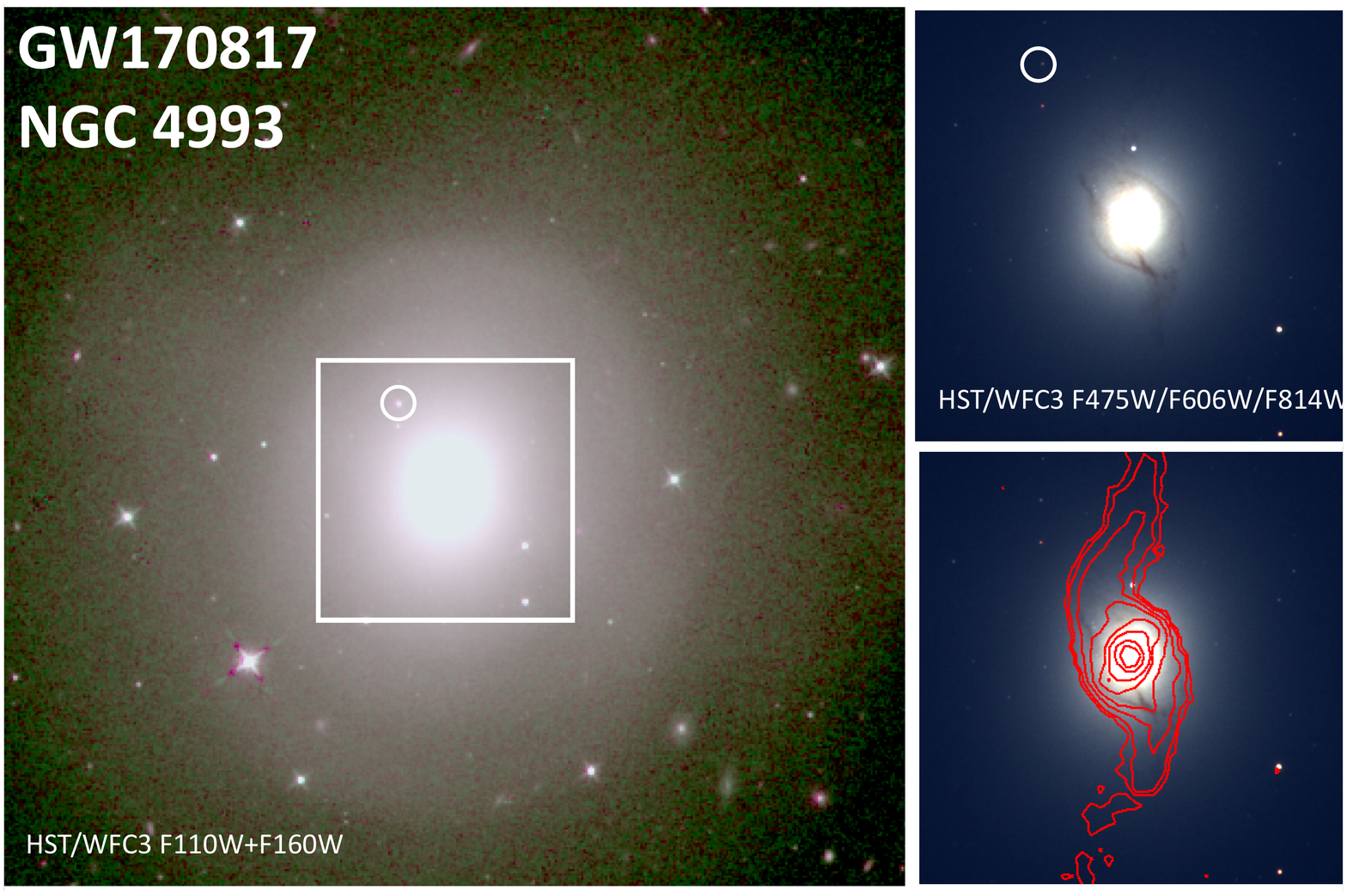}
\caption{Imaging of the host galaxy of GW170817 with {\em HST}. 
The left hand panel shows the galaxy observed in the IR with F110W and F160W, where the counterpart is marked with a circle. The top right panel shows the zoomed in region observed with WFC3/UVIS, demonstrated the presence of strong dust lanes in the inner regions. The lower panel shows the same image, but with the MUSE contours in the [N\,{\sc ii}] line 
superimposed, showing the strong spiral features that only appear in the emission lines. Some of these features appear to trace the dust lanes.}
\label{finder}
\end{figure*}

Face-on spiral shell-like features can be seen extending to large radii ($\sim 1$\arcmin\ 
or $12$ kpc) in our
{\em HST} imaging. Emission line fits were attempted within the MUSE cube throughout the host after subtraction of the underlying stellar continuum. Ionised gas can be seen in spiral arms with a relatively strong ($\sim0.8$~kpc) bar (see Figures~\ref{finder} and \ref{musen2}). Notably, these spiral features seen in nebular emission lines appear with a high ratio of minor to major-axes, suggesting an almost edge on alignment. The velocity structure in these edge-on spiral arms extends to $\sim 220$ km s$^{-1}$ within the central kpc of the galaxy, while the stellar components appear to be dominated by much lower velocities ($\sim 100$~~km~s$^{-1}$, Figure~\ref{musesp}). Furthermore, the large scale spiral arms/shells are almost circular (and hence face on), while the spiral features seen in emission lines are near edge-on. The decoupled dynamics of the gas and stars suggest a relatively recent merger, and that the galaxy has not yet relaxed. This would also be consistent with the presence of dust lanes in a quiescent galaxy \citep[e.g.][]{kaviraj12,shabala12}, as seen in our {\em HST} optical imaging (see Figure~\ref{finder}). This recent merger would provide the natural fuel to power nuclear accretion.

The presence of extended emission lines could imply star formation in the host galaxy, however, the emission line ratios are difficult to explain with photoionization from young stars. We find [N{\sc ii}] $\lambda 6583$/H$\alpha$ (and [O{\sc iii}] $\lambda 5007$/H$\beta$, where it could be measured) ratios of $\sim 1$ (Figure~\ref{musen2}).
Such ratios are typical of those seen in AGN and LINERs (low-ionisation nuclear emission-line regions). 
However, the relative weakness of any AGN, combined the large spatial extent of these lines and their spiral structure is very different to typical extended emission line regions seen around some AGN. We therefore conclude that AGN power is not responsible for their creation.
Instead, these features are similar to the so-called LIER regions (non-nuclear LINER regions) that have been found in other early type galaxies without AGN activity \citep{sarzi10,singh13}. These may be driven by intermediate-aged stellar populations (e.g. hot post-AGB stars) or shocks. Since these emission lines are not excited by young stars associated with recent star formation, they cannot be used to reliably derive a gas-phase metallicity for the galaxy.

\section{At the transient location}

Our imaging shows the
transient is located at position offset 8.92\arcsec N and 5.18" E
of the host galaxy centroid, with a total projected offset of 10.31 $\pm$ 0.01\arcsec, corresponding to 1.96 kpc offset at a 40 Mpc distance\footnote{This provides a lower limit on the true offset since we cannot measure the source location relative to its host along our line on sight}. Our observations provide constraints on the immediate environment of the transient, and on any underlying source.

\subsection{In emission}
Any source underlying the transient position is of significant interest, since it could indicate either a young stellar cluster (and hence young progenitor), or a globular cluster in which the NS-NS binary may have formed dynamically \citep{grindlay06}.
At the location of the transient in the F606W imaging obtained prior to GW170817 \citep[see also][]{foley17} we place a 2$\sigma$ upper limit on any point source by subtracting an isophotal fit
to the smooth light of the galaxy, and then performing aperture photometry on the galaxy subtracted image. The resulting point source limit is F606W$>26.4$ ($2 \sigma$), 
corresponding to an absolute magnitude of $M_V > -6.7$ (note at 40 Mpc globular clusters should appear point like as the HST PSF is $\sim 20$ pc, indeed point sources in the images around the galaxy have colours consistent with globular clusters associated with NGC 4993). 
This is well down the globular cluster luminosity function and only $\sim 30$\% of Milky Way globular clusters would evade detection at
this limit \citep{harris}. However, clusters fainter than this limit only contain
$\sim 5$\% of the stellar mass in globular clusters. We consider the implications of this further in section 5.

There is no detection of the counterpart in F275W images taken on 25 Aug 2017 and 28 Aug 2017 \citep[see][]{troja}. Hence, in addition to constraining the counterpart they can also place limits on underlying young populations. Combining two epochs of F275W observations (exposure time 1240s) places a limit on the UV luminosity of any source of F275W$>26.0$ (2$\sigma$, AB), corresponding to an absolute magnitude limit of F275W(AB)$\gtrsim-7$. 
This absolute magnitude limit 
is comparable to the UV absolute magnitudes of massive O-stars or Wolf-Rayet stars, and suggests little underlying star formation. Indeed, combined with the optical limit from pre-imaging, this indicates that any underlying young ($10^7 - 10^8$ yr) star cluster could have a mass of only a few thousand solar masses \citep[based on the BPASS models of][]{eldridge17}. 

Given the lack of any point source we can characterise the location of the transient relative to the host galaxy light, defining the fraction of
surface brightness contained in pixels of equal or lower surface brightness to the pixel hosting the transient, the so-called $F_{light}$ parameter \citep{fruchter06}. This statistic is complicated in this case by the extended low surface brightness features, and the presence of foreground stars projected onto the galaxy. Masking these stars, and considering light within a large ($1.2\arcmin$ radius) aperture suggests that $F_\mathrm{light} \approx 0.6$. This is at the upper end of those seen in short GRBs \citep{fong}, although it should be noted that in SGRBs at higher redshift, cosmological surface brightness dimming could result in the omission of low surface brightness features in the comparison samples. The loss of light at low surface brightness (compared to the transient position) would lower the value of $F_{light}$.
Alternatively, one can consider the host normalized offset (the offset of the transient from its host nucleus in units of the effective radius), which in this instance is also $\sim 0.6$, smaller than the typical SGRB offsets \citep{fong}. While these values are consistent with the distributions seen for short GRBs, the location of GW170817 within its host galaxy is comparable to the most centrally-concentrated 10\% of short GRBs.

\subsection{In absorption}
We searched for narrow absorption features in the transient spectrum obtained with MUSE, but did not find any convincing examples. The transient's features are too broad to be useful for velocity measurements, especially since
their identification is uncertain. Hence the 
velocity of the source relative to the host galaxy cannot be directly determined. 

\begin{figure*}[h]
    \centering
    \includegraphics[width=0.9\linewidth,angle=0]{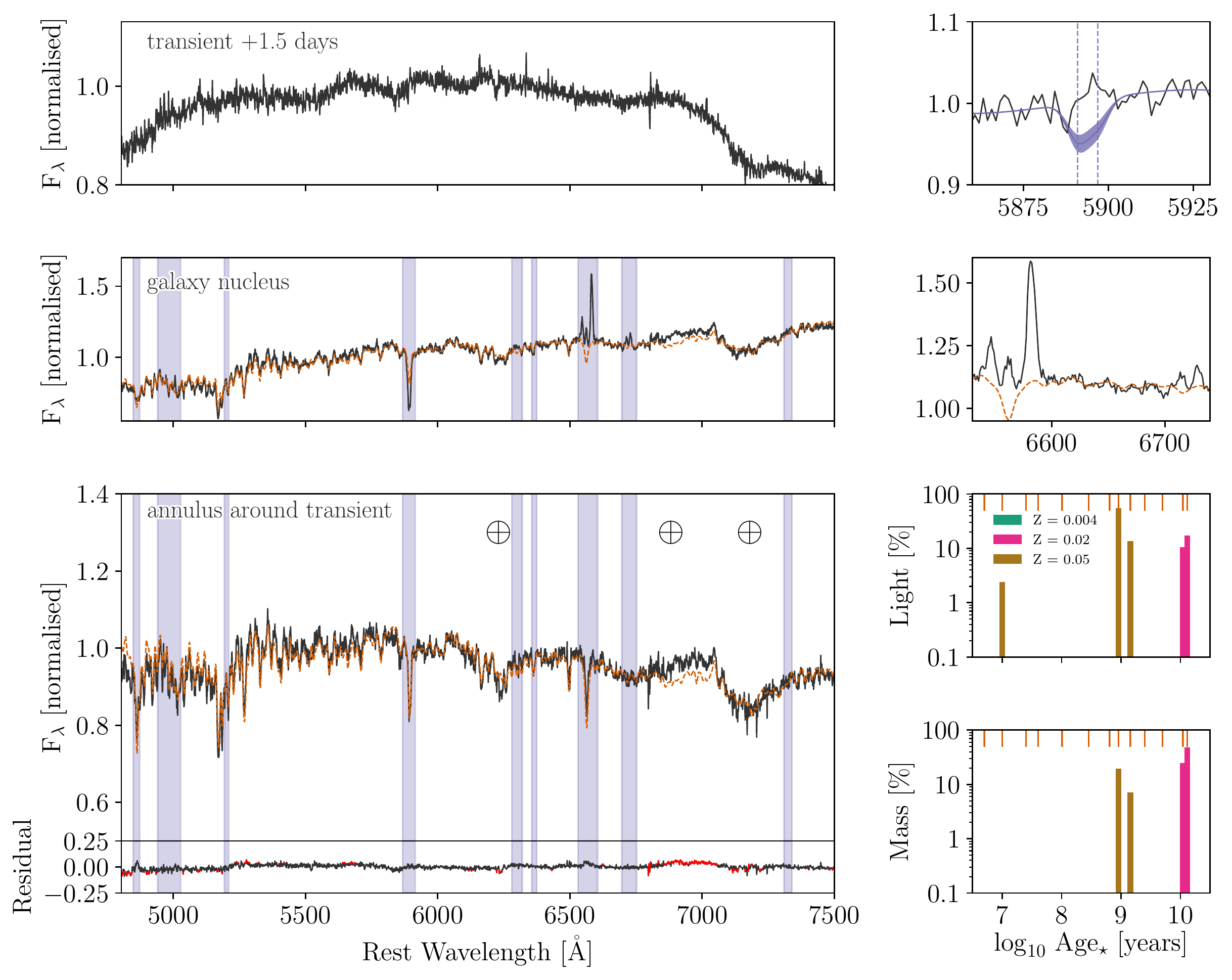}
\caption{{\em Bottom:} Stellar population results in an annulus 1.5--2$\arcsec$ around the transient. The extracted MUSE spectrum (black) with model continuum fit (orange, dashed) and residuals (lower panel). Wavelengths clipped are indicated in red in the residual panel and largely coincide with regions of telluric features (indicated on fit panel). Shaded regions are masked from the fit. The light and mass contribution of the stellar population models by age (abscissa) and metallicity by mass fraction (colour) to the total fit are shown on the right. Orange tick marks (top axis) indicate the ages of all models used in the fit. The small contribution of the 10~Myr stellar population is not robust to varying the model set used. We consider it an upper estimate of any young stellar population contribution. The best fit extinction is $A_V = 0.22$~mag. {\em Middle:} Stellar population fit for the nucleus of NGC~4993. The conspicuous emission lines of H$\alpha$, [N{\sc ii}] and [S{\sc ii}] are shown in a zoomed in panel on the right. {\em Top:} A 0.4$\arcsec$ radius aperture extracted at the transient location after subtracting the annulus spectrum (scaled by the ratio of  the area of the aperture and annulus). The right panel shows a zoom in around Na {\sc i} $\lambda\lambda5890, 5896$. Overlaid on the transient spectrum is the Na{\sc i} absorption expected based on the extinction derived in the annulus, using the equivalent width relations of \citet[][see text]{poznanski12} with a 1$\sigma$ shaded region. The absorption was simply modeled as two gaussians with $\sigma = \sigma_\star$(annulus) plus a low order polynomial to fit the continuum. All spectra are normalised to the flux in the range 5590--5680~\AA{}}
\label{musefit}
\end{figure*}

\begin{figure*}[h]
    \centering
    \includegraphics[width=\linewidth,angle=0]{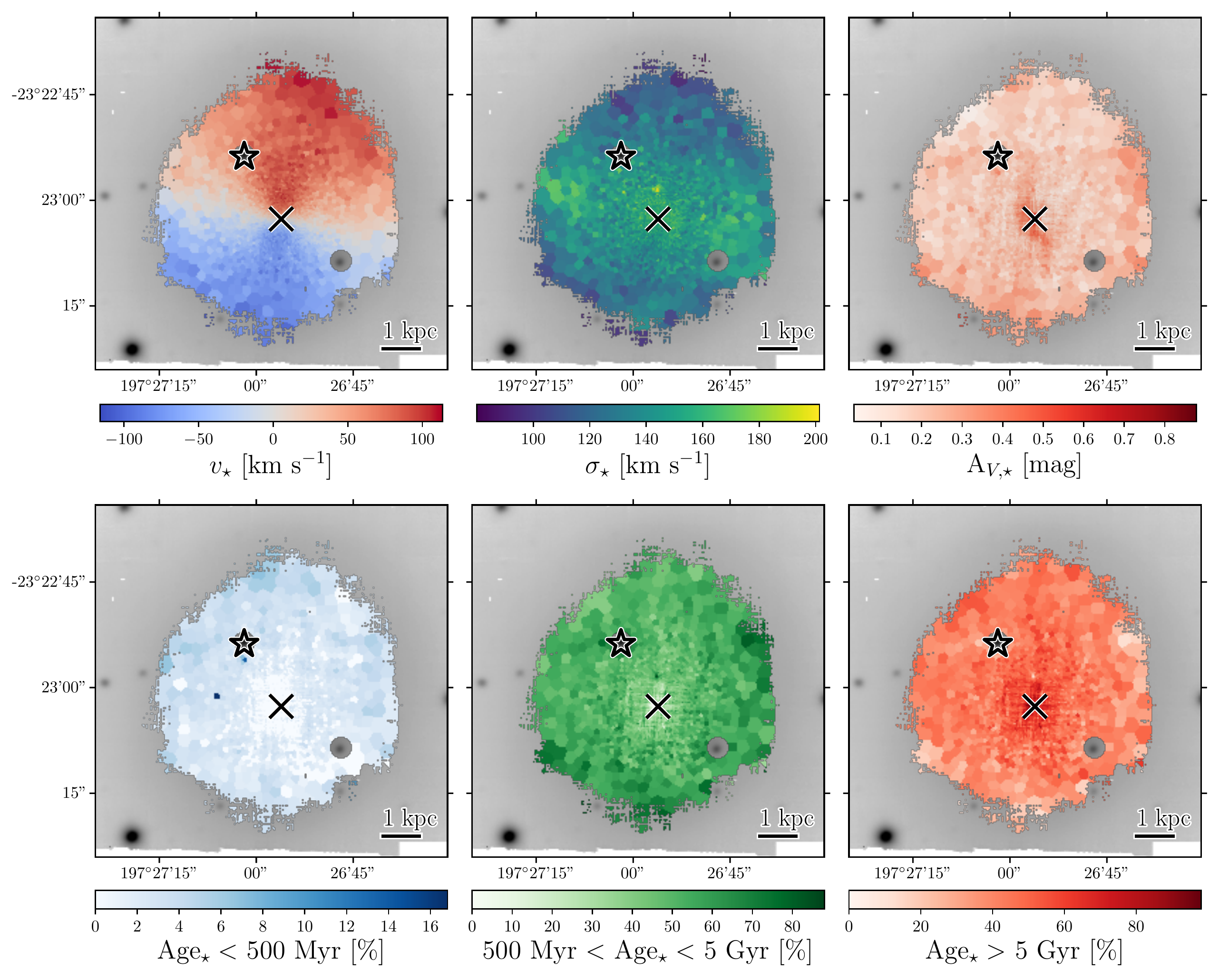}
\caption{Voronoi binned stellar population property maps of NGC~4993 from our stellar continuum fitting. {\em Top row:} stellar velocity offset relative to the galaxy nucleus {\em (left)}, stellar velocity dispersion {\em (middle)} and visual extinction {\em (right)}. {\em Bottom row:} The flux contribution to the best fit continuum model for young {\em (left)}, intermediate {\em (middle)} and old {\em (right)} stellar populations; the age divisions are indicated on the respective colour bars. The galaxy nucleus and transient locations are indicated by a $\times$ and star symbols, respectively. North is up, East is left and the linear scale at a distance of 40~Mpc is shown. Maps are overlaid on a collapsed view of the MUSE datacube. Masks (grey circles) have been applied around the transient and a foreground star.}
\label{musesp}
\end{figure*}

\begin{figure*}[h]
    \centering
    \includegraphics[width=0.6\linewidth,angle=0]{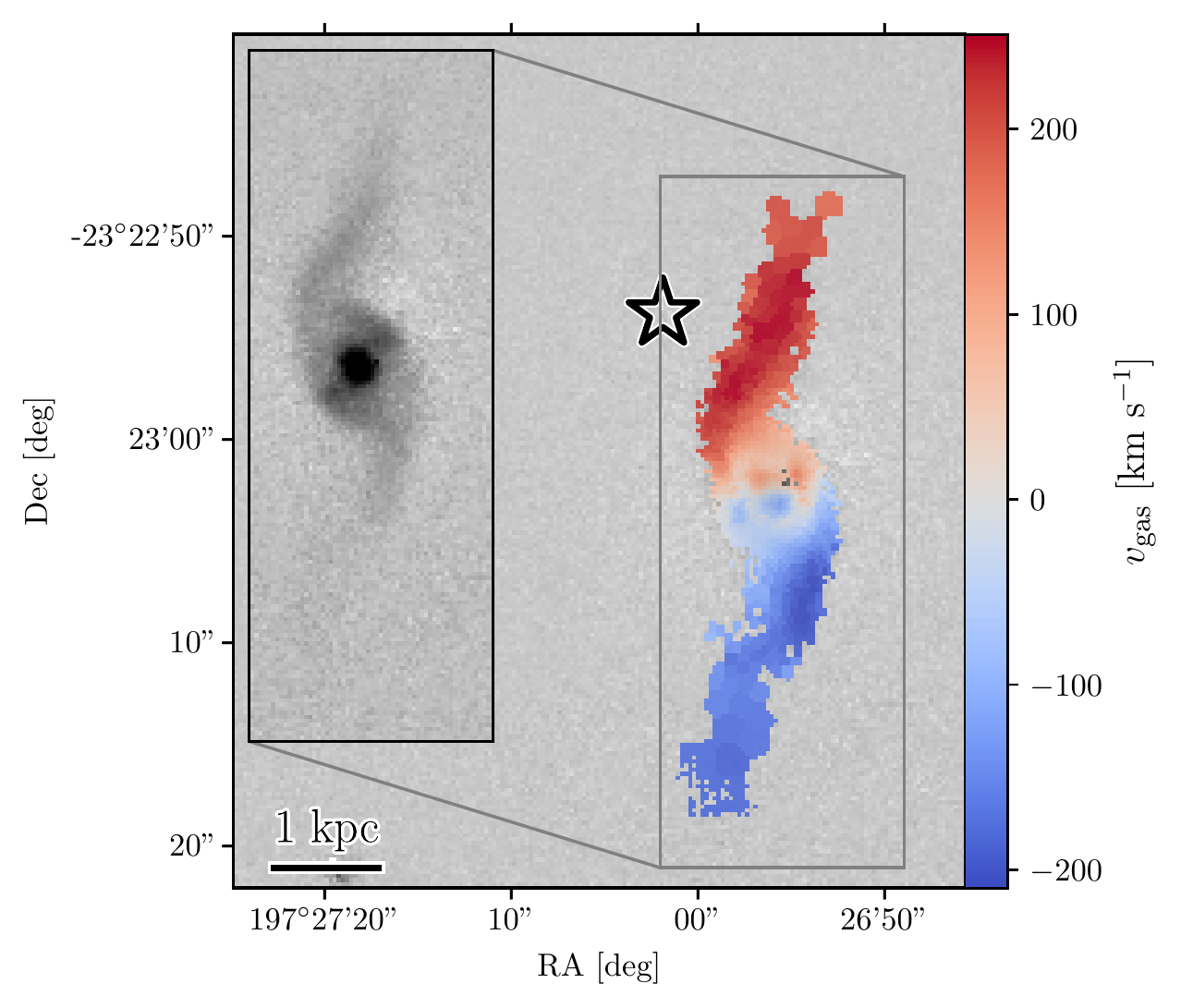}
    \includegraphics[width=0.6\linewidth,angle=0]{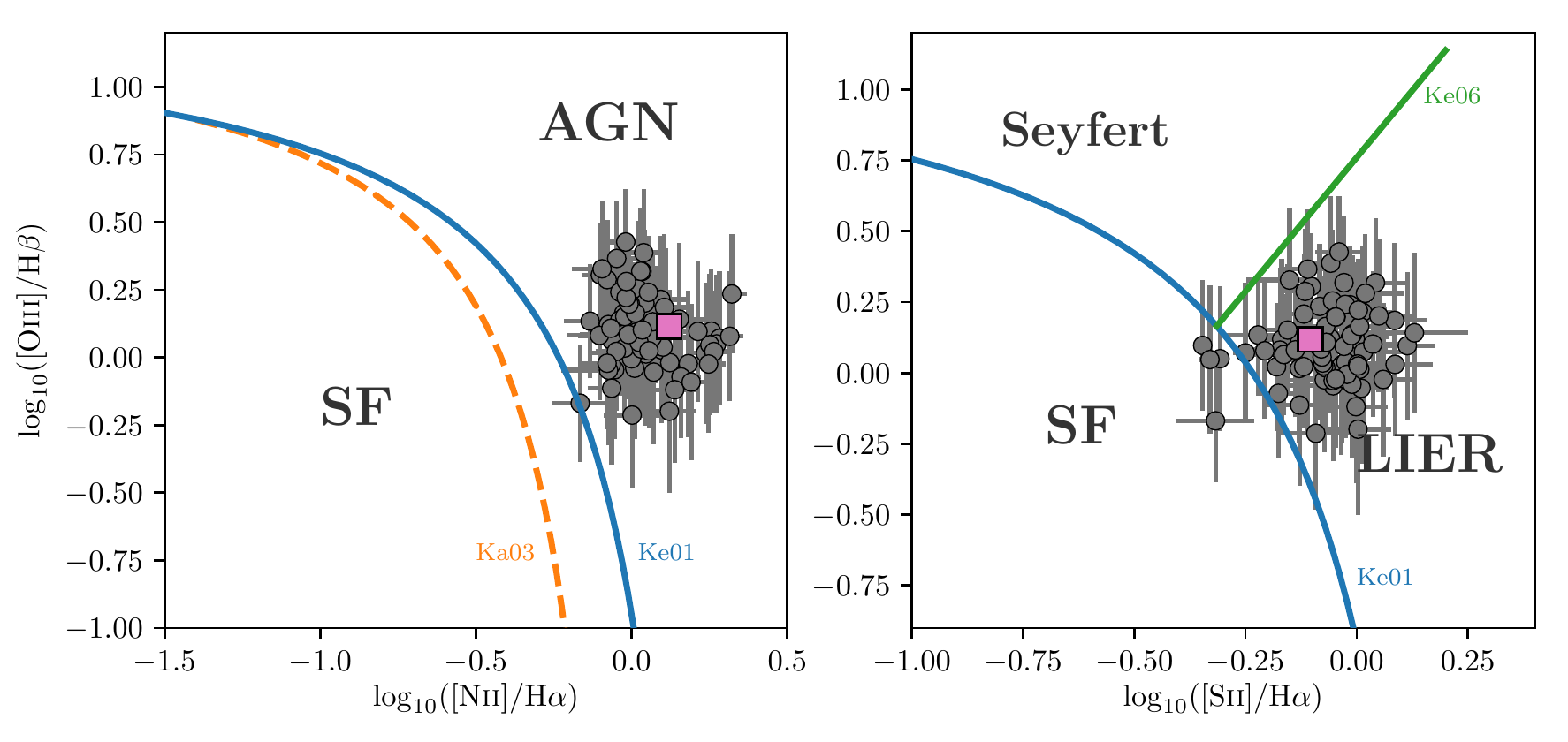}
\caption{{\em Top:} Velocity map of ionised gas in NGC~4993. Velocities are the offsets of [N\,{\sc ii}] $\lambda$6583 emission. The location of the transient is indicated with a star symbol. Inset shows an [N\,{\sc ii}] $\lambda$6583 narrow-band image constructed from the MUSE data cube. The motion of the ``arm'' like features is in the radial velocity plane, implying they are edge on.  {\em Bottom:} Emission line ratios in NGC~4993. Each point represents a spaxel bin where all the diagnostic lines are detected as a signal-to-noise ratio of $> 2$. Dividing lines of \citet[][pure star-formation; orange dashed]{kauffmann03}, \citet[][theoretical star-formation limit; blue]{kewley01} and \citet[][Seyfert vs. LIER; green]{kewley06} are shown. The ionised gas regions in NGC~4993 are inconsistent with being driven by star formation (SF) and are better described as extended LIERs \citep[see][]{sarzi10, singh13}. The few bins formally within theoretical star-formation limits show velocity dispersions $\sigma_\textrm{gas} > 100$~km~s$^{-1}$, inconsistent with typical H{\sc ii} regions that show tens~km~s$^{-1}$.   The position of the integrated emission lines (summing all bins) is shown by a square on each plot.}
\label{musen2}
\end{figure*}

Notably, the region around the counterpart shows evidence for modest extinction of $E(B-V) = 0.07$~mag based on the stellar population fits (see Figure~\ref{musefit}),
assuming an $R_V=3.1$ extinction law. If the 
extinction directly underlying the transient position is the same, then in principle this should be visible as Na{\sc i} D absorption in the counterpart spectrum, the equivalent width of which can be calculated from established relations \citep{poznanski12}. Indeed, this doublet is seen
in the sole short GRB spectrum to show absorption features in its afterglow \citep{adup14}. The expected lines for $E(B-V) = 0.07$~mag are shown in Figure~\ref{musefit} where we have subtracted a scaled version of the annulus spectrum in order to correct the transient spectrum for the Na{\sc i} D absorption that arises from a stellar origin within the transient aperture,  with the shaded region demonstrating the $1\sigma$ scatter in this relation. There is no evidence for such absorption. The absence of ISM absorption in the counterpart spectrum
could imply that the transient emission may lie in front of the stellar population (or gas) in the galaxy. This may be because it is naturally located within the halo population, or alternatively could be indicative that the progenitor received a kick which has placed it outside the bulk of the stellar population. Ultimately a re-observation of the explosion site, once the transient has faded, will allow us to investigate this further by directly measuring the extinction to the stellar population under the transient position. However, since the extinction measured in the annulus around the transient is typical of extinction across the galaxy (see top right panel of Fig 3), our value seems reasonable.

\section{Comparison with expectations from binary evolution}

The location of a binary neutron star merger depends on its initial location, the delay
time between the formation of the second NS and the merger, the kick given to the binary at the time of NS formation, and the galactic potential in which the system moves. 

Within the Milky Way, the majority of binary neutron stars are formed in the field through isolated binary evolution, while a single example exists in a globular cluster, likely formed via dynamical interactions \citep{anderson90}, and similar interactions may produce a significant fraction of extragalactic systems \citep{grindlay06}.

There is no evidence in our data supporting the dynamical formation scenario. The closest point-like sources in the stellar field of the galaxy are approximately 2.5\arcsec\ from the transient location, corresponding to an offset of $\sim 500$ pc (a lower limit owing to projection effects).  The absence of a globular cluster at the transient location strongly disfavors a merger within a globular cluster. Dynamical formation involves a sequence of 3-body (2+1) interactions that leave more massive stars within tighter binaries \citep{davies95}. The rate of these interactions scales roughly as $L^{3/2}/r^{5/2}$ \citep{bregman06}, indicating that massive, core collapsed systems disproportionately dominate the interaction rate \citep{davies95b}. In this sense the faintest 30\% of globular clusters contain 5\% of the mass in globular clusters, but probably have $\ll 5\%$ of the interactions, indicating a very low probability of an underlying cluster creating the binary that formed GW170817. 
However, two caveats apply to this. 
Firstly, if a significant population of black holes remains in the cluster, the more massive black holes will substitute into compact binaries during 2+1 interactions and BH-BH binaries would be formed at the expense of NS-NS binaries \citep[e.g.,][]{rodriguez16}.  Secondly, interactions can eject the binaries given the low escape velocities of globular clusters.  With modest velocities and long delay times these may be indistinguishable from field systems based on their locations, although it may be possible to distinguish field and dynamically formed binaries based on their intrinsic properties (masses, spins, etc.) measured from the gravitational waves themselves \citep[e.g.,][]{stevenson17,zevin17,farr17}.

The absence of any significant young population within the galaxy, or of young stars underneath the burst position, suggests that the progenitor of GW170817 was likely old ($>10^9$ yr). This old age is consistent with the ages of Milky Way double neutron star systems \citep[e.g.,][]{tauris17}. However, rapidly merging systems are observationally selected against (because they merge quickly, the time when they are detectable is short), and some population synthesis suggests that a significant fraction could have very short delay times if the progenitor of the second-born neutron star either enters a second common envelope phase after the helium main sequence \citep{dewi03} or stably transfers a significant amount of mass which is then lost from the binary \citep[e.g.][]{bel06,2016MNRAS.462.3302E,vdh17}.

The delay times for some of these models $\sim 10^6$ years or less \citep{bel06}, and so binaries could not travel far from their birth location. The absence of any underlying cluster or point source close to the transient position can therefore offer information on the likelihood of a very short delay progenitor. Firstly, it is unlikely that a low mass cluster would be present close to the transient position without other signs of star formation more widely distributed in the host galaxy. Secondly, such a low mass cluster would be unlikely to form many massive stars. For a typical initial mass function \citep[e.g.][]{kroupa03}, a 1000 M$_{\odot}$ cluster would be expected to form $\sim 10$ stars capable of forming supernovae. Since $\lesssim 10$ merging binary neutron stars are expected to form per million solar masses of star formation \citep{abadie10}, the probability of obtaining such a binary from a single low mass star forming region is very small. 

Observations of a single object cannot provide a delay time distribution for a binary population, but do demonstrate on this occasion that the delay time was large.

The question of the kick velocity is even more difficult to determine. The space velocity of a binary neutron star relative to the initial velocity of its centre of mass in the galactic potential is set by the kick imparted at each supernova. Such a kick has two contributing factors, the natal kick to the neutron star, and the mass loss kick from the binary. The majority of binaries are disrupted during one of the supernova events; only those for which the mass loss is small, and the natal kick fortuitous in direction, are likely to survive. For this reason it is likely that the binary neutron star population has slower velocities than isolated pulsars, and there is evidence to suggest this is true \citep{dewi03}.

\begin{figure*}[h]
    \centering
    \includegraphics[width=0.8\linewidth,angle=0]{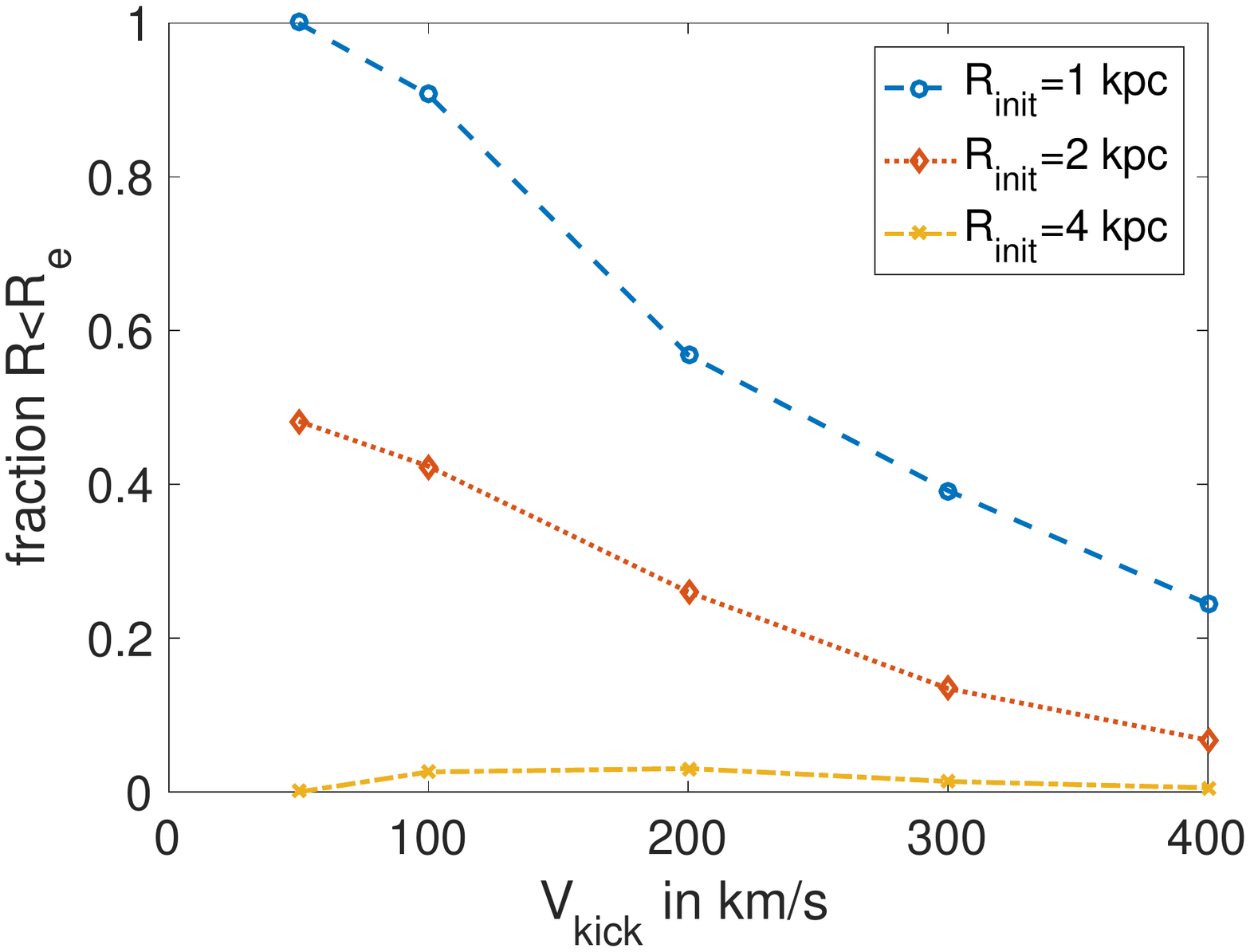}
    \vspace{-1.7in}
\caption{The fraction of time that a binary with a given kick velocity magnitude spends within $R_e = 2$ kpc of the galactic center.  Binaries are assumed to form on circular orbits around the galactic center at initial distances of $1$ kpc (blue), $2$ kpc (orange) or $4$ kpc (yellow) from the center.   For formation at $4$ kpc from the center and kicks of $400$ km/s, $7\%$ of binaries are unbound; these are not included in the plot.  All other simulated binaries are bound, and the plot is reflective of the merger location if the delay time between double compact object formation and gravitational-wave driven merger is longer than the dynamical timescale.}
\label{fractionRe}
\end{figure*}

Once kicked, binaries move in the gravitational potential of their host galaxy. For high-velocity kicks they may be ejected completely, or move on highly elliptical orbits.
For weak kicks, they may orbit much closer to the galaxy, and may oscillate around the radius at which they were formed.  We quantify this in Figure~\ref{fractionRe}, where we show binaries kicked isotropically at different velocities from circular orbits at three initial radii within the host galaxy. Once kicked these binaries are assumed to move in the gravitational potential equal to the Milky Way potential model of \cite{irrgang13} with the disc removed. As a bulge dominated lenticular galaxy there is little, if any disc component in NGC 4993, and this model provides a good description of the rotational velocities seen in Fig 4. For long delay times ($\gg$ orbital period of the binary around the galaxy), the probability of merging beyond a given radius (for example, the effective radius) is equal to the fraction of time the binary spends beyond this radius.

As indicated in Figure \ref{fractionRe}, a strongly kicked binary with a long delay time is likely to be found at radii outside the effective radius, whatever its initial location.  Indeed, the presence of the counterpart of GW170817 relatively close to the centre of the host would favour a small kick if the system is old, as our stellar population analysis suggests.  As with the issue of delay times, a single event provides little information as to the properties of the population, but it is worth noting that the proximity of GW170817 to its host would place it amongst the most centrally concentrated $\sim$10\% of short GRBs\citep[e.g.,][]{fong}. This may suggest that the kick in GW170817 was unusually small, that it was directed towards us (minimizing the projected offset), or that a binary in a more extended orbit happened to merge when passing relatively close to the core of its host galaxy.

\section{Summary and conclusions}
We have presented comprehensive imaging and integral-field spectroscopy of the host galaxy and local environment of the first electromagnetic counterpart to a gravitational wave source. These observations provide a unique view of the regions around this event, and its properties are consistent with those seen in the population of short-duration GRB hosts.
We find a highly-inclined ionised gas disk that is kinematically-decoupled from the stellar velocity field, as well as extended face-on arm/shell features in the stellar light profile. These indicate the galaxy has undergone a major merger relatively recently. We find that $\sim 20$\% of the galaxy, by mass, is $\sim$~1~Gyr old, perhaps as a result of this merger, while most of the remaining mass is $> 5$ Gyr old. There is minimal contribution (if any) from a young stellar population ($\ll 1\%$ of the mass), implying an old ($\gtrsim$~Gyr) progenitor. The absence of absorption features in the counterpart spectrum and moderate extinction of the stellar population in the vicinity of the transient source offer tentative evidence that it lies on the near side of the galaxy, either by chance or due to a kick in our direction. 

Galaxy demographics and population synthesis have previously been used to argue for the origin of short duration GRBs in compact object mergers. Since this scenario now seems secure the direction of inference can now be reversed, and the properties and locations of short GRBs and gravitational wave sources can be used to pinpoint the details of extreme stellar evolution that lead to the formation of compact object binaries.

\begin{table}[ht]
\begin{center}
\begin{tabular}{lll}
\hline
Instrument   & Band  & Magnitude (AB) \\
\hline
HST/WFC3     & F275W & 19.87 $\pm$ 0.15 \\
VLT/VIMOS    & $U$   & 14.96 $\pm$ 0.01 \\
VLT/VIMOS    & $R$   & 12.13 $\pm$ 0.01 \\
VLT/VIMOS    & $z$   & 11.60 $\pm$ 0.01 \\
HST/WFC3     & F110W & 11.27 $\pm$ 0.01 \\
HST/WFC3     & F160W & 10.94 $\pm$ 0.01 \\
VLT/HAWK-I   & $K$   & 11.50 $\pm$ 0.01 \\
Spitzer/IRAC & 3.6 $\mu$m & 11.81 $\pm$ 0.02 \\
Spitzer/IRAC & 4.5 $\mu$m & 12.34 $\pm$ 0.02 \\
\hline
\end{tabular}
\end{center}
\caption{Photometry of NGC 4993. Photometry has been measured in 1\arcmin apertures centred on the host galaxy, and bright foreground stars have been masked from the image (except for F275W where a smaller aperture was used due both to the centrally concentrated nature of the UV emission, and the windowed WFC3 FOV). The errors given are statistical only. Given the uncertainy introduced by the masking, and possible low surface brightness features it is reasonable to assume systematics of $\sim$ 0.1 mag. Magnitudes are not corrected for foreground extinction.}
\label{obs}
\end{table}%

\acknowledgments
Acknowledgments.
We thank the referee for a prompt and highly constructive report that improved the content and clarity of the manuscript. We also thank the editor, Fred Rasio for helpful comments.
Based on observations made with ESO Telescopes at the La Silla Paranal Observatory under programme ID 099.D-0668 (Levan), and
on observations made with the NASA/ESA Hubble Space Telescope, obtained from the data archive at the Space Telescope Science Institute. STScI is operated by the Association of Universities for Research in Astronomy, Inc. under NASA contract NAS 5-26555. These observations
are associated with programs GO 14771 (Tanvir), GO 14804 (Levan), and GO 14850 (Troja). We thank the staff at ESO and STScI for their excellent support of these observations. AJL acknowledges that this project has received funding from the European Research Council (ERC) under the European Union's Horizon 2020 research and innovation programme (grant agreement no 725246)
AJL, DS, JDL acknowledge support from STFC via grant ST/P000495/1. NRT, KW, PTO, JLO, SR acknowledge support from STFC.
JH was supported by a VILLUM FONDEN Investigator grant (project number 16599). AdUP, CT, ZC, and DAK acknowledge support from the Spanish project AYA 2014-58381-P.  ZC also acknowledges support from the Juan de la Cierva Incorporaci\'on fellowship IJCI-2014-21669, and DAK from Juan de la Cierva Incorporaci\'on fellowship IJCI-2015-26153. MI was supported by the NRFK grant, No. 2017R1A3A3001362. ET acknowledges support from grants GO718062A and HSTG014850001A. SR has been supported by the Swedish Research Council (VR) under grant number 2016- 03657\_3, by the Swedish National Space Board under grant number Dnr. 107/16 and by the research environment grant ``Gravitational Radiation and Electromagnetic Astrophysical Transients (GREAT)" funded by the Swedish Research council (VR) under Dnr 2016-06012. PAE acknowledges UKSA support.

\facilities{Hubble Space Telescope, Very Large Telescope} 
\software{Numpy, PyRAF, {\sc astropy} \citep{astropy}, {\sc starlight} \citep{starlight}, {\sc Reflex} \citep{reflex}}

\end{document}